%% file: make_astro.tex
\documentclass{mpe_report}

\usepackage{psfig,graphicx,epsfig}
\usepackage{color}
\usepackage{amsmath,amssymb,epic,eepic,array}

\begin{document}

\pagenumbering{arabic}
\setcounter{page}{13}

\renewcommand{\FirstPageOfPaper }{ 13}\renewcommand{\LastPageOfPaper }{ 15}\include{./mpe_report_hui}                  \clearpage

\end{document}

%% file: mpe_report_hui.tex
\def\PSR{PSR J2124$-$3358}

\title{Discovery of an X-ray Nebula associated with PSR J2124$-$3358}
\author{C. Y. Hui \and W. Becker}  
\institute{Max--Planck--Institut f\"ur extraterrestrische Physik,
 Giessenbachstra{\ss}e, 85740 Garching, Germany}
\maketitle

\begin{abstract}
We report the discovery of an X-ray nebula associated with the 
nearby millisecond pulsar \PSR. This is the first time that 
extended emission from a solitary millisecond pulsar is detected. 
The emission extends from the pulsar to the northwest by $\sim0.5$ 
arcmin. The spectrum of the nebular emission can be modeled by a 
power law spectrum with photon index of $2.2\pm0.4$. This is inline 
with the emission being originated from accelerated particles in the 
post shock flow. 
\end{abstract}

\section{Introduction}
\begin{figure}
\psfig{file=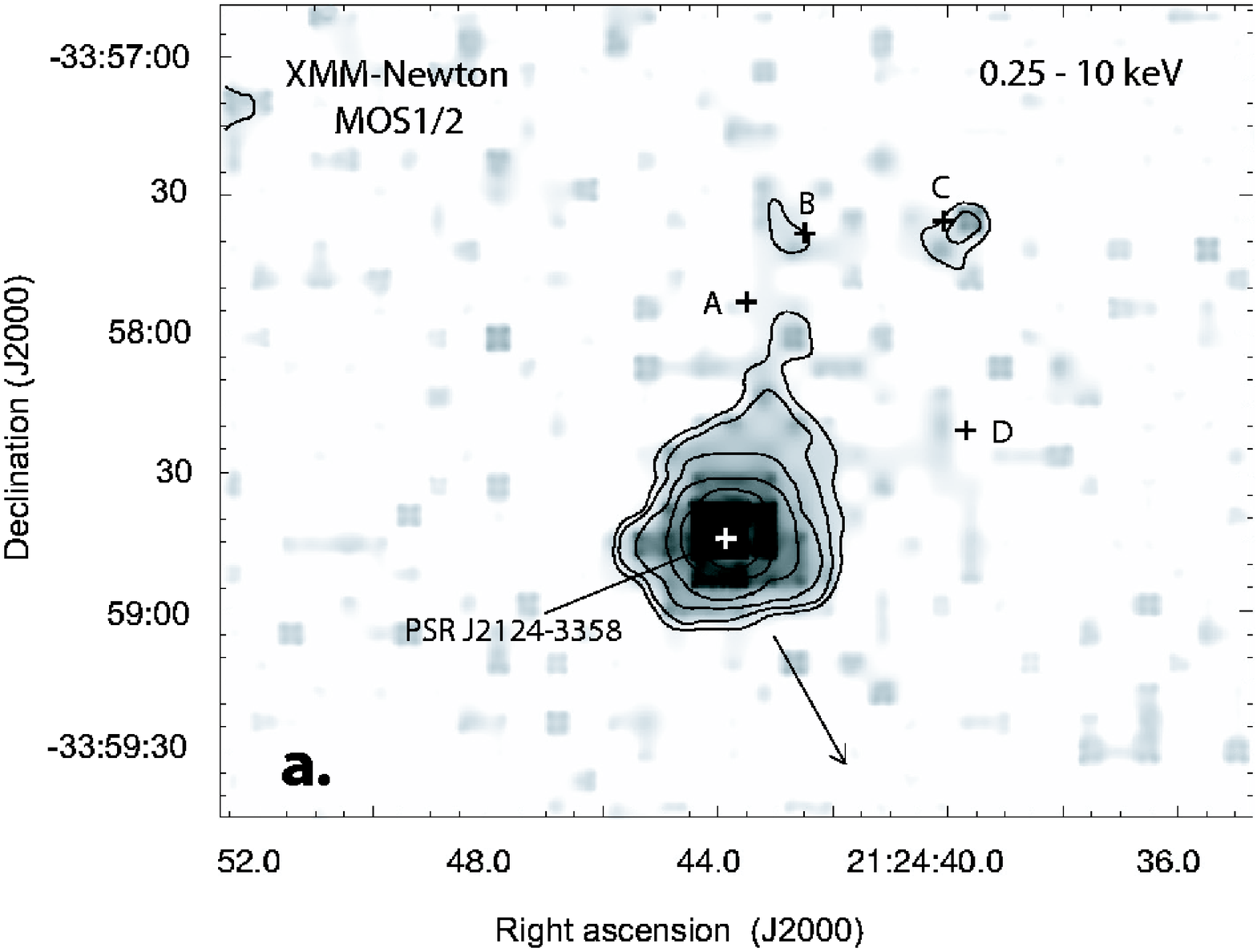,width=10cm,clip=}
\psfig{file=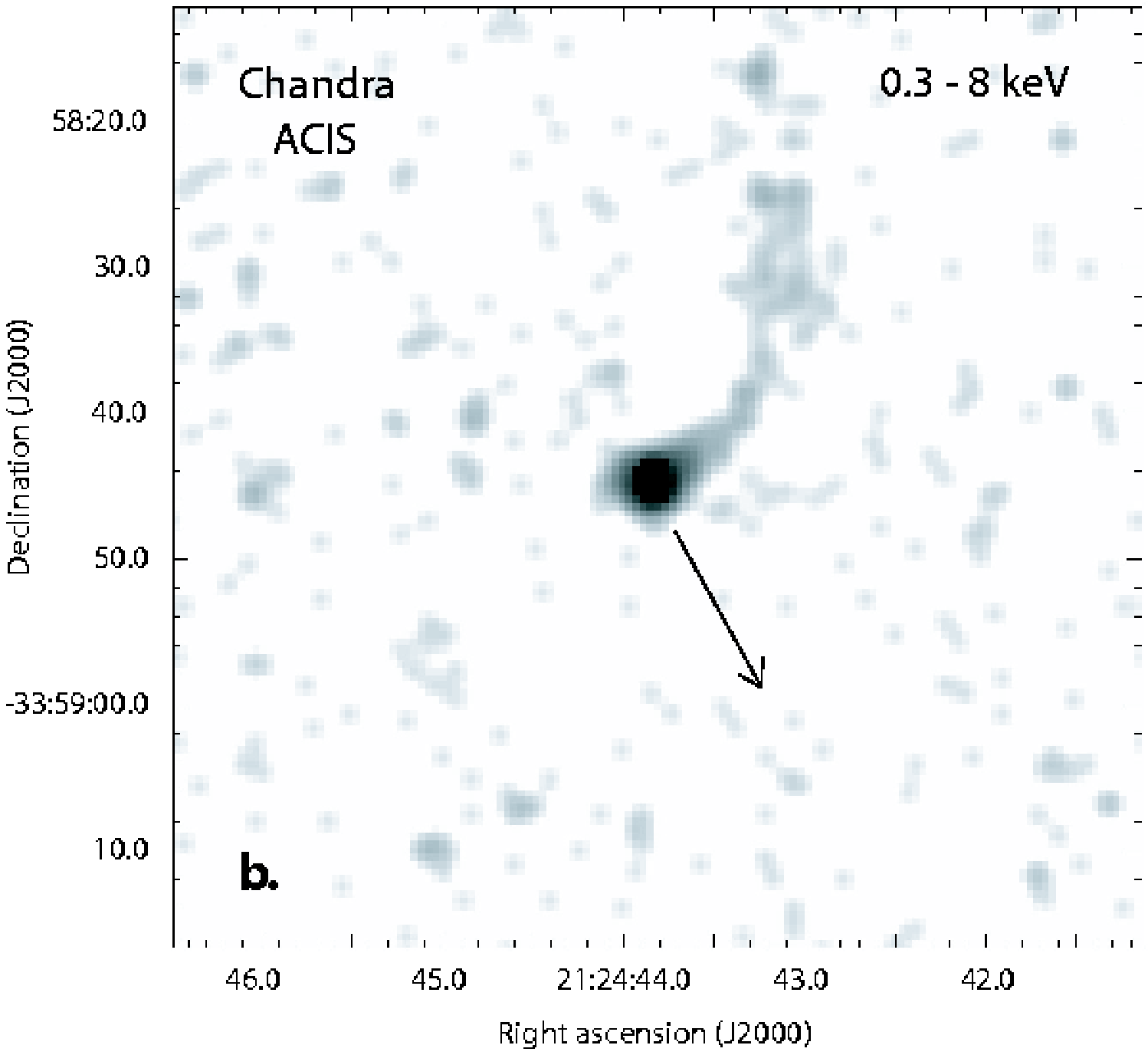,width=10cm,clip=}
\caption{a) XMM-Newton MOS1/2 image of \PSR\ with overlaid contours. The pulsar proper
motion is indicated by an arrow.  The position of bright stars located in the 1.5 arcmin
neighborhood is indicated.  b) \PSR\ as seen by the ACIS detector aboard Chandra.}
\vspace{-0.7cm}
\end{figure}

Rotation-powered pulsars are well-known to be the objects tapping their rotational energy
into pulsed emission. However, it is now generally believed that a much larger fraction of
the rotational energy leaves the pulsars' magnetosphere in the form of magnetized wind. 
When the relativistic wind particles interact with the interstellar
medium, synchrotron emission, which is characterized by a power-law spectrum
(Chevalier 2000) is radiated. 

If a pulsar moves across the surrounding medium with a velocity exceeding the speed of
sound for the medium, a bow shock can be formed. The structure of a bow shock can consist of
two parts (Gaensler 2005). The forward shock, resulting from collisional excitation, is
expected to produce H$\alpha$ emission. Also, the termination shock is produced by the
pressure difference between regions ahead of and behind the pulsar's velocity vector. The
relativistic wind particles will be accelerated in the termination shock and hence generate
synchrotron emission from radio to X-ray. The morphology of the X-ray termination shock is
generally cometary.

Recently, we have conveyed searches for diffuse X-ray emission around a group of millisecond pulsars. 
The pulsars are \PSR, J0437$-$4715, J0030$+$0451 and J1024$-$0719 which have comparable spin 
parameters (Hui \& Becker 2006). The period and period derivative of this group have ranges of 
4.87$-$5.76 ms and (1.0$-$1.87)$\times 10^{-20}$ s s$^{-1}$ respectively. In this work, we have 
discovered an elongated structure associated with \PSR\ (Figure 1). 

PSR J2124-3358 was discovered by Bailes et al.~(1997) during the
Parkes 436 MHz survey of the southern sky. The pulsar has a rotation period of $P=4.93$ ms
and a proper motion corrected period derivative of $\dot{P}=1.33\times10^{-20}$ s s$^{-1}$.
These spin parameters imply a characteristic age of $5.86\times10^{9}$ yrs and a dipole
surface magnetic field of $2.60\times10^{8}$ G (Manchester et al.~2005). The radio dispersion
measure gives a distance of about 250 kpc. It is found to have a space velocity of
$\sim58$ km/s (Manchester et al.~2005). Gaensler, Jones \& Stappers (2002) discovered an H$\alpha$-emitting 
bow shock nebula around PSR J2124-3358. This bow shock is very broad and highly
asymmetric about the direction of the pulsar's proper motion. 

\section{Data Analysis and Results}

\PSR\ has been observed by XMM-Newton in 2002 April 14$-$15 with an effective exposure time of $\sim40$ 
ks. A vignetting corrected image of the field of PSR J2124-3358 as seen by the XMM-Newton's
MOS1/2 CCDs is shown in Figure 1a. The binning factor in this image is 6 arcsec.
Adaptive smoothing with a Gaussian kernel of $\sigma<4$ pixels has been applied to
the image. X-ray contours are calculated and overlaid on the image. The contour lines
are at the levels of $(4.2, 5.2, 7.6, 13, 28, 63)\times10^{-6}$ cts s$^{-1}$ arcsec$^{-2}$.
It can be seen that the X-ray source which is coincident with the pulsar position
has an asymmetric source structure of $\sim 0.5$ arcmin extent, with its orientation
to the northwest. Systematic effects which could cause an adequate distortion of
the instrument's point spread function (PSF) are not known for XMM-Newton. We are
therefore prompted to interpret this elongated structure in terms of a pulsar X-ray
trail. The signal-to-noise of this elongated feature in the XMM-Newton data 
is $\sim 4$ in the energy range $0.25-5$ keV.

We have investigated a possible contribution to the diffuse X-ray emission by nearby
stars. To do so, we investigated the Digitized Sky Survey data (DSS) for the sky region around
\PSR. There are four field stars, which are labeled as A, B, C and D in Figure 1a, 
in the 1.5 arcmin neighborhood of \PSR. None of them is found to match the position of the diffuse
elongated X-ray structure. It can be seen in
Figure 1a that the positions of  stars B and C coincide with two faint X-ray sources
which are disconnected with the trail emission of \PSR, though.

The detection is which supported by the Chandra ACIS-S3 observation took place on 2004 
December 19-20 for an exposure time of $\sim 30$ ks. An
image made from this data with 0.5 arcsec binning and adaptive smoothing applied (using a
Gaussian kernel with $\sigma < 1.5$ pixel) is shown in Figure 1b. Arc-like diffuse emission
which is within the pulsar's H$_\alpha$ nebula is clearly detected. 
Since the PSF width of XMM-Newton is about 10 times that of Chandra, it blurred most of the
detailed structure seen in the Chandra data. However, it should be noted that the
overall direction of the feature in the Chandra image is consistent with the
orientation of the trail detected by XMM-Newton. The signal-to-noise of this feature in the Chandra data 
is $\sim 5$ in the energy range of $0.3-8$ keV.

\begin{figure}
\centerline{\psfig{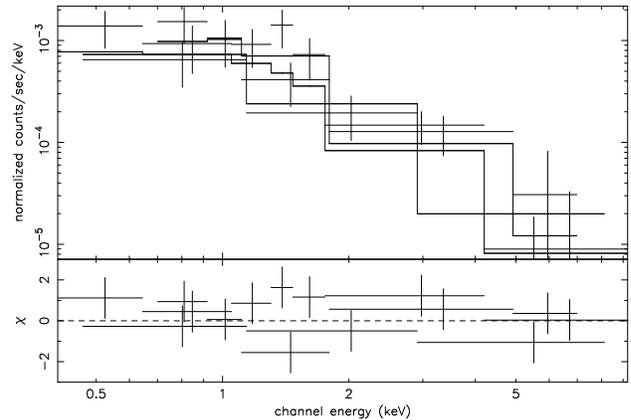}}
\caption{Energy spectrum of the X-ray trail of \PSR\, as observed with
MOS1/2 and ACIS-S3 detectors and simultaneously fitted to an absorbed power-law model
({\it upper panel}) and contribution to the $\chi^{2}$ fit statistic
({\it lower panel}).}
\vspace{-0.4cm}
\end{figure}

For the spectral analysis, we extracted the energy spectrum within a $30\times35$
arcsec box from the XMM-Newton MOS1/2 data. 
Using a box rather than a circular selection region allows to avoid the
emission from the pulsar and excludes any potential contamination from the field
stars B and C. However, we estimate that the wing of the XMM-Newton PSF centered at
the pulsar position still contributes $\sim 20\%$ to the total counts inside the box.
The background  spectrum was extracted from a source free region near to \PSR. In total,
92 and 67 source counts are available from the trail in the MOS1 and MOS2 detectors,
respectively. Response files were computed by using the XMMSAS tasks RMFGEN and ARFGEN.
The spectra were dynamically binned so as to have at least 10 counts per bin.

In the Chandra data we selected the energy spectrum of the diffuse emission from a box
of size $10\times 20$ arcsec. Owing to the narrow PSF of Chandra the contamination
of pulsar emission in this box is negligible. The background spectrum was extracted
from a low count region near to the diffuse feature. In total 46 source counts are
contributed from the Chandra data. Response files were computed by using tools MKRMF
and MKARF of CIAO. The spectrum was binned to have at least 8 counts per bin.
The degradation of quantum efficiency of ACIS was corrected by applying the XSPEC model ACISABS.

We hypothesize that the diffused emission should be originated from the interaction
of pulsar wind and the ISM. Synchrotron radiation from the ultra-relativistic electrons
is generally believed to be the emission mechanism of the pulsar wind nebula, which then
is characterized by a power-law spectrum. With a view to test this hypothesis, we fitted
an absorbed power-law model to the nebular spectrum with XSPEC 11.3.1 in the $0.25-10$ keV
energy range. With a column density of $5\times10^{20}$ cm$^{-2}$ as obtained from spectral
fits to the pulsar emission, we found that the model describes the observed spectrum fairly
well ($\chi^{2}_{\nu}=0.79$ for 26 D.O.F.). The photon index is $\alpha=2.2\pm0.4$ and the
normalization at 1 keV is $(2.94\pm0.48)\times10^{-6}$ photons keV$^{-1}$ cm$^{-2}$ s$^{-1}$
($1-\sigma$ error for 1 parameter in interest). In view of the low  photon statistic we tested
for a possible dependence of the fitted model parameters against the background spectrum.
All deviations found in repeating the fits with different background spectra were within the
$1-\sigma$ confidence interval quoted above. The unabsorbed fluxes and luminosities deduced
for the best fit model parameters and the energy ranges 0.1$-$2.4 keV and 0.5$-$10 keV are
$f_{X}=1.8\times10^{-14}$ ergs s$^{-1}$ cm$^{-2}$, $L_{X}=1.3\times10^{29}$ ergs s$^{-1}$
and $f_{X}=1.2\times10^{-14}$ ergs s$^{-1}$  cm$^{-2}$, $L_{X}=8.9\times10^{28}$ ergs s$^{-1}$,
respectively. The best fitting spectral model is displayed in Figure 2.

\section{Discussions}

Adopting the dispersion measure based distance of $\sim$ 250 pc, the X-ray trail 
has a length of $l\sim1.1 \times10^{17}$ cm. For the pulsar's proper motion velocity
of 58 km s$^{-1}$ (Manchester et al.~2005), the timescale, $t_{\rm flow}$, for the passage of
the pulsar over the length of its X-ray trail is estimated to be $\sim600$ yrs.  According
to the discussion in Becker et al.~(2005) on the trail emission of PSR B1929+10 we estimate
the magnetic field in the shocked region by assuming $t_{\rm flow}$ to be comparable to the electron lifetime
of the synchrotron emission. This yields $\sim30\mu$G for the inferred magnetic field strength in the
emitting region. The magnetic field strength in the ISM is estimated to be $\sim2-6\mu$G (cf.~Beck et al.~2003
and references therein). Taking into account that the magnetic field in the termination shock might be
compressed (e.g.~Kennel \& Coroniti 1984), our estimation is approximately consistent if the compression
factor is $\sim 7$.

Following Becker et al.~(2005), we applied a simple one zone model (Chevalier 2000) to estimate the spectral
behavior and the X-ray luminosity of the nebular emission.  The X-ray luminosity and spectral index depend
on the inequality between the characteristic observed frequency $\nu^{\rm obs}_{X}$ and the electron synchrotron
cooling frequency $\nu_{\rm c}$ which is estimated to be $1.6\times10^{17}$ Hz. Since in general
$\nu^{\rm obs}_{X}>\nu_{\rm c}$, we concluded that the emission is in a fast cooling regime. Electrons with the
energy distribution, $N(\gamma)\propto\gamma^{-p}$, are able to radiate their energy in the trail with
photon index $\alpha=(p+2)/2$. The index $p$ due to shock acceleration typically lies between 2 and 3
(cf. Cheng, Taam, \& Wang 2004 and references therein). Taking $p=2.35$ yields $\alpha^{\rm th}\simeq2.2$
which is in accordance with the result from the observed value $\alpha^{\rm obs}=2.2\pm0.4$. Assuming the
energy equipartition between the electron and proton (Cheng, Taam, \& Wang 2004), we take the fractional
energy density of electron $\epsilon_{\rm e}$ to be $\sim0.5$ and the fractional energy density of the magnetic
field $\epsilon_{B}$ to be $\sim0.01$. Assuming a number density of ISM to be 1 cm$^{-3}$, the distance of
the shock from the pulsar is estimated to be $\sim3.6\times10^{16}$ cm. With these estimates, the calculated
luminosity, $\nu L_{\nu}$, is given as $\sim10^{29}$ ergs s$^{-1}$ which is well consistent with the observed
values of $1.3\times10^{29}$ ergs s$^{-1}$ (0.1-2.4 keV) and $8.9\times10^{28}$ ergs s$^{-1}$ (0.5-10 keV).

Although the general properties of the X-ray trail in \PSR\, are not in contradiction with properties
observed in other pulsars there are still ambiguities which are not completely resolved.
First, one should notice that the trail is misaligned with the direction of the pulsar's proper
motion. As reported by Gaensler, Jones, \& Stappers (2002), the head of the $H_\alpha$ bow shock is found
to be highly asymmetric about the pulsar's velocity vector with the apparent nebular symmetry axis deviated
from the velocity vector by $\sim30^{\circ}$. Even though the misalignment of the X-ray trail seems
to agree with the asymmetry of the $H_\alpha$ nebula, deeper observations by XMM-Newton and Chandra
are required in order to constrain the physical properties of this interesting nebula in higher detail.

\vskip 0.4cm

\begin{acknowledgements}
We gratefully acknowledge the support by the WE-Heraeus foundation.
\end{acknowledgements}
   
